\documentclass{cjaa}                   
\usepackage{graphicx}                   
\input{epsf.sty}                        
\input{psfig.sty}                       

\renewcommand{\baselinestretch}{1.1}

\newcommand{\relmodv}{$\langle|V|\rangle/S$}
\newcommand{\relv}{$\langle V\rangle/S$}

\begin{document}

\title{
Circular Polarization in Pulsar Integrated Profiles: Updates~$^*$
\footnotetext{$*$ Supported by the National Natural Science
Foundation of China.} }

\volnopage{Vol.\ 6 (2006), No.\ 2,~ 237--246}
   \setcounter{page}{237}

   \author{Xiao-Peng You
      \inst{}
   \and Jin-lin Han
      \inst{}
      }

   \institute{National Astronomical Observatories, Chinese Academy of Sciences,
             Beijing 100012; {\it xpyou@bao.ac.cn}\\
\vs\no        {\small Received 2005 June 29; accepted 2005
December  5 }}

\abstract{We update the systematic studies of circular
polarization in integrated pulse profiles by Han et al (1998).
Data of circular polarization profiles are compiled. Sense
reversals can occur in core or cone components, or near the
intersection between components. The correlation between the sense
of circular polarization and the sense of position angle variation
for conal-double pulsars is confirmed with a much large database.
Circular polarization of some pulsars has clear changes with
frequency. Circular polarization of millisecond pulsars is
marginally different from that of normal pulsars.
 }

\authorrunning{X. P. You \& J. L. Han }            
\titlerunning{ Circular Polarization in Pulsar Integrated Profiles: Updates }  

 \vspace{-8mm}  \maketitle

   \maketitle

\section{Introduction}
\label{sect:intro}

Polarization properties of pulsars are very important for the
understanding of the geometry and emission mechanism of pulsars.
Generally, the degree of circular polarization is low. Many
pulsars show sense reversal in their circular polarization near
the middle of the pulse. The sense reversals sometimes are
associated with the orthogonal polarization modes (\cite{C78};
\cite{S84a}). In some pulsars, the circular polarization keeps the
same sense through the whole profile. Two obvious types of
circular polarization are identified by \cite{RR90}, namely:
$antisymmetric$, where the circular polarization changes sense
near the center of the profile, and $symmetric$, where the
circular polarization remains the same sense through the whole
profile. \cite{Han98} collected the published polarization
profiles and reviewed the characteristics of circular polarization
in pulsar integrated profiles, discovered a correlation between
the sense of circular polarization and the sense of position angle
(PA) variation for conal-double pulsars, and rebutted the
correlation between the sense reversal of circular polarization
near the core components and the sense of PA.

There are two possible origins of circular polarization of
pulsars: either intrinsic to the emission properties and dependent
on the emission mechanism, or generated by propagation effects.
For example, \cite{ML04} discussed possible circular polarization
induced by intrinsically relativistic effects of pulsar plasma.
\cite{melrose03} reviewed the properties of intrinsic circular
polarization and circular polarization due to cyclotron
absorption, and presented a plausible explanation of circular
polarization in terms of propagation effects in an inhomogeneous
birefringent plasma. \cite{LP99} considered that the rotation of
the magnetosphere gives rise to wave mode coupling in the
polarization-limiting region, which can result in circular
polarization in linearly polarized normal waves.

A large sample of normal pulsars and millisecond pulsars has been
observed for polarization (\cite{GL98}; \cite{S99}; \cite{W99};
\cite{W04}; \cite{H06}), especially at multiple frequencies. The
data have increased by a factor of about three over that in
\cite{Han98}. So, it is the time to update the database of pulsar
circular polarization and recheck the conclusions of that paper.

\section{Dataset}

Polarization profiles of pulsars are collected and cataloged if
the circular polarization has a good signal-to-noise ratio.
Circular polarization is defined observationally by the Stokes
parameter, $V= I_L-I_R$. The rotational sense of $V$, the
percentage (=\relv, where $S$ is the mean total flux density), and
absolute circular polarization percentage, \relmodv, the variation
of PA, and observation frequency are all included in Table~1 (This
is only part of Table~1. For the full Table~\ref{tb:big_tab} see
http://www.chjaa.org/2006v1n2/ for electronic version), which has
the same format as table~A1 in \cite{Han98}.

\begin{table}[h!!]

\centering \begin{minipage}{110mm} \caption{A Summary of Pulsar
Circular Polarization Observations}
\label{tb:big_tab}\end{minipage}

\scriptsize
\begin{tabular}{lllrrclclp{30mm}}
\hline\noalign{\smallskip}
 PSR J & PSR B & $V$ &\multicolumn{1}{c}\relmodv &  \multicolumn{1}{c}\relv & $\sigma$ &
 PA & \multicolumn{1}{r}{Freq.}  & Ref. & Comments \\
   $$  &       &     & (\%)      & (\%)  & (\%)     &    &   (MHz)                    &      &   \\
\hline\noalign{\smallskip}
0030$+$0454   &           &$  --      $&  &$   $&  &dec &  433 &L00  &MSP. $--$ for main comp \\
0034$-$0534   &           &$  --      $&18&$   $&  &xx  &  410 &S99  &PA swing not clear \\
0034$-$0721   & 0031$-$07   &$  +       $&10&$  5$& 1&dec &  234 &G98  & \\
              &           &$  +       $&  &$   $&  &dec &  268 &R83  &$+V$ in 2nd half \\
              &           &$  +       $&  &$   $&  &dec &  328 &S05  &$+V$ in 2nd half \\
              &           &$  -       $& 6&$ -6$& 0&inc &  410 &G98  &$-V$ in 1st half \\
              &           &$  +       $& 5&$  4$& 0&i+d &  606 &G98  &$+V$ in 2nd half \\
              &           &$  +       $&  &$   $&  &dec & 4850 &S05  &$+V$ in 2nd half \\
0040$+$5716   & 0037$+$56   &$  -       $&18&$-17$& 1&inc &  610 &G98  &s/n good\\
0045$-$7319   &           &$  -       $&27&$-17$& 7&xx  &  661 &C01  &strong cp \\
0048$+$3412   & 0045$+$33   &$  -       $&  &$   $&  &dec &  430
&W04  &PA not very clear. low linear
polarization\\\noalign{\smallskip} \hline
\end{tabular}
\end{table}

\section{Main properties of circular polarization}

\subsection{Sense
Reversal of Circular Polarization}

\subsubsection{Sense reversals associated with core components}

Using a sample of 25 pulsars, \cite{RR90} found that change of
circular polarization from left hand (positive) to right hand
(negative) is associated with decreasing PA, and that from right
hand to left hand is associated with increasing PA. Gould (1994)
and \cite{Han98} found many contrary examples, which leads
\cite{Han98} to conclude that no correlation exists between the
sense of the sign change of circular polarization and the sense of
variation of PA.

Here we use a very large sample of pulsar data and confirm the
conclusion of non-correlation. Table~\ref{tb:core_rev} lists all
pulsars with sense reversal of circular polarization in the core
component, and 19 pulsars in the first and fourth part of
Table~\ref{tb:core_rev} support the existence of the correlation,
but 20 pulsars in the second and third part do not.

\begin{table}

\centering
\begin{minipage}{135mm}
 \caption{Sense Reversals of Circular Polarization
Associated with Core Components} \label{tb:core_rev}
\end{minipage}

\scriptsize

\begin{minipage}{145mm}

\begin{minipage}[t]{70mm}
\tabcolsep 1mm

\begin{tabular}{lclcr}
\hline\noalign{\smallskip}
PSR Name     &$V=LH-RH$& PA  & \multicolumn{1}{c}{Freq.} & Ref.  \\
              &         &     &  (MHz)                    &       \\
              \hline\noalign{\smallskip}
J0454$+$5543  &  $+/-$  & dec &  610                      & G98   \\
J0809$-$4753  &  $+/-$  & dec &  1335                     & H06   \\
J1001$-$5507$^a$&  $+/-$  & dec &  1351                     & H06   \\
J1239$+$2453  &  $+/-$  & dec &  1418                     & W99   \\
J1509$+$5531  &  $+/-$  & dec &  610                      & G98   \\
J1534$-$5334  &  $+/-$  & dec &  1612                     & Ma80  \\
J1740$+$1311  &  $+/-$  & dec &  1418                     & W99   \\
J1801$-$0357$^b$&  $+/-$  & dec &  661                      & M98   \\
J1823$+$0550  &  $+/-$  & dec &  1408                     & G98   \\
J1900$-$2600  &  $+/-$  & dec &  1408                     & G98   \\
J1901$+$0331  &  $+/-$  & dec &  1418                     & W99   \\
J1903$-$0632  &  $+/-$  & dec &  610                      & G98   \\
J1946$-$2913  &  $+/-$  & dec &  1327                     & H06   \\
J2048$-$1616  &  $+/-$  & dec &  1420                     & LM88  \\
J2113$+$4644  &  $+/-$  & dec &  610                      & G98
\\[1mm]
J0826$+$2637m &  $+/-$  & inc &  1400                     & R89   \\
J1512$-$5759  &  $+/-$  & inc &  1319                     & H06   \\
J1600$-$3053$^c$ &  $+/-$  & inc &  1373                     & O04   \\
J1604$-$4909  &  $+/-$  & inc &  658                      & M98 \\
J1733$-$2228$^b$ &  $+/-$  & inc &  610                   & G98   \\
\hline\\
\end{tabular}
\end{minipage}~~
\begin{minipage}[t]{70mm}
\tabcolsep 1mm
\begin{tabular}{lclcr}
\hline\noalign{\smallskip}
PSR Name     &$V=LH-RH$& PA  & \multicolumn{1}{c}{Freq.} & Ref.  \\
              &         &     &  (MHz)                    &       \\
              \hline\noalign{\smallskip}
J1910$-$0309  &  $+/-$  & inc &  1408                     & G98   \\
J1926$+$0431  &  $+/-$  & inc &  1418                     & W99   \\
J2004$+$3137  &  $+/-$  & inc &  1400                      & R89   \\
J2006$-$0807  &  $+/-$  & inc &  1408                     & G98
\\[1mm]
J0332$+$5434  &  $-/+$  & dec &  1408                     & G98   \\
J0437$-$4715  &  $-/+$  & dec &  1512                     & N97  \\
J0944$-$1354  &  $-/+$  & dec &  409                      & LM88  \\
J1326$-$5859  &  $-/+$  & dec &  955                      & v97   \\
J1456$-$6843  &  $-/+$  & dec &  649                      & Mc78  \\
J1527$-$5552  &  $-/+$  & dec &  658                      & M98   \\
J1537$+$1155  &  $-/+$  & dec &  430                      & A96   \\
J1544$-$5308  &  $-/+$  & dec &  658                      & M98   \\
J1752$-$2806  &  $-/+$  & dec &  1408                     & G98   \\
J1852$-$2610  &  $-/+$  & dec &  434                      & M98   \\
J1909$+$0254$^b$  &  $-/+$  & dec &  610                      &
G98   \\[1mm]
J0452$-$1759  &  $-/+$  & inc &  408                      & LM88  \\
J1703$-$3241  &  $-/+$  & inc &  950                      & v97   \\
J2144$-$3933  &  $-/+$  & inc &  659                      & M98   \\
J2325$+$6316$^c$ &  $-/+$  & inc &  1642                     & G98
\\[2mm]
\hline\\
\end{tabular}
\end{minipage}
 { $^a$ New high resolution observation show decreasing PA.
Old data are confused by orthogonal polarization modes. \\
$^b$ PA not clear.\\
$^c$ Not so sure if sense reversal happens in core.\\
}
\end{minipage}

\centering 

\begin{minipage}{100mm}
\fns \caption{Sense Reversals not Associated with Core Components}
\label{tb:other_rev}\end{minipage}

\setlength{\tabcolsep}{0.9mm}

\scriptsize
\begin{tabular}{lclcll}
\hline\noalign{\smallskip}
PSR Name       &$V=LH-RH$& PA  & \multicolumn{1}{c}{Freq.} & Ref. & Comments \\
              &         &     &  (MHz)                    &      &  \\
\hline\noalign{\smallskip}
J1651$-$4246  &$  +/-  $& dec & 1349        & H06  & intersection between two comp\\
J1807$-$0847  &$  +/-  $& dec & 1408        & G98  & cone\\
J1857$+$0943m &$  +/-  $& dec & 1400        & S86  & intersection between two comp\\
J1907$+$4002  &$  +/-  $& dec & 1408        & G98  & intersection between two comp?\\
J2324$-$6054  &$  +/-  $& dec & 1335        & H06  & intersection between two comp\\[1mm]
J0152$-$1637  &$  +/-  $& inc & 660         & Q95  & intersection between two comp\\
J0612$+$3721  &$  +/-  $& inc & 610         & G98  & associated with orthogonal polarization modes\\
J0653$+$8051  &$  +/-  $& inc & 610         & G98  & cone? \\
J0738$-$4042  &$  +/-  $& inc & 1351        & H06  & associated with orthogonal polarization modes \\
J0837$+$0610  &$  +/-  $& inc & 800         & S84b & leading cone  \\
J1913$-$0440  &$  +/-  $& inc & 408         & G98  & associated with orthogonal polarization modes \\
J2022$+$2854  &$  +/-  $& inc & 800         & S84b & leading cone  \\
J2053$-$7200  &$  +/-  $& inc & 658         & M98  & intersection between two comp\\
              &$  -/+  $& inc & 1440        & Q95  & intersection between two comp\\
J2145$-$0750  &$  +/-  $& inc & 610         & S99  & cone \\
J2326$+$6113  &$  +/-  $& inc & 1408        & G98  & intersection between two comp?\\
J1708$-$3426  &$  +/-  $& ??  & 1329        & H06  & intersection between two comp\\[1mm]
J0133$-$6957  &$  -/+  $& dec & 658         & M98  & intersection between two comp\\
J1041$-$1942  &$  -/+  $& dec & 1642        & G98  & leading cone  \\
J1045$-$4509  &$  -/+  $& dec & 1373        & O04  & cone? \\
J1559$-$4438  &$  -/+  $& dec & 1490        & M98  & intersection between trailing two comp\\
J1614$+$0737  &$  -/+  $& dec & 1418        & W99  & intersection between two comp\\
J1705$-$1906m &$  -/+  $& dec & 1642        & G98  & intersection between two comp\\
J1751$-$4657  &$  -/+  $& dec & 434         & M98  & leading cone  \\
J1916$+$0951  &$  -/+  $& dec & 610         & G98  & leading cone  \\
J1935$+$1616  &$  -/+  $& dec & 1408        & G98  & intersection between two comp\\[1mm]
J0255$-$5304  &$  -/+  $& inc & 1359        & H06  & second cone \\
J0502$+$4654  &$  -/+  $& inc & 1408        & G98  & intersection between two comp\\
J0601$-$0527  &$  -/+  $& inc & 408         & G98  & intersection between two comp\\
J0941$-$5244  &$  -/+  $& inc & 1319        & H06  & intersection between two comp\\
J1224$-$6407  &$  -/+  $& inc & 1319        & H06  & intersection between two leading comp  \\
J1328$-$4357  &$  -/+  $& inc & 435         & M98  & intersection between two leading comp  \\
J1921$+$2153  &$  -/+  $& inc & 1418        & W99  & cone\\
J1602$-$5100  &$  -/+  $& ??  & 950         & v97  & intersection between two leading comp  \\
\hline\noalign{\smallskip}
 \end{tabular}
\end{table}

\subsubsection{Sense reversals outside core}

Many sense reversals of circular polarization are detected outside
of the center region of the profile, thus not associated with core
components but with cone components or near the conjunction of
components. Table~\ref{tb:other_rev} lists such pulsars with sense
reversals in the other part of the pulse profile.

\subsubsection{Sense reversals associated with orthogonal polarization modes}

We checked possible association of sense reversal of circular
polarization with orthogonal polarization modes of the
polarization angle. Among 81 pulsars with sense reversals in $V$
with clear PA variation curves, about 31 show the association. For
example, the PA jumps about $90^\circ$ seen in PSRs J1900$-$2600,
J0601$-$0527 and J0437$-$4715 at almost all the observed
frequencies (\cite{M98}; \cite{GL98}; \cite{N97}), near the phase
of a sense transition of circular polarization. A few pulsars show
two sense reversals across the profile, as shown in Figure~1 for
PSR J2037$+$1942 which has sense reversals associated with the
peaks of two components. The orthogonal polarization modes occur
in the first component (\cite{W99}). The polarization curve thus
does not have a good $S-$shape.

\begin{figure}[h!!]

  \centering
\includegraphics[width=80mm]{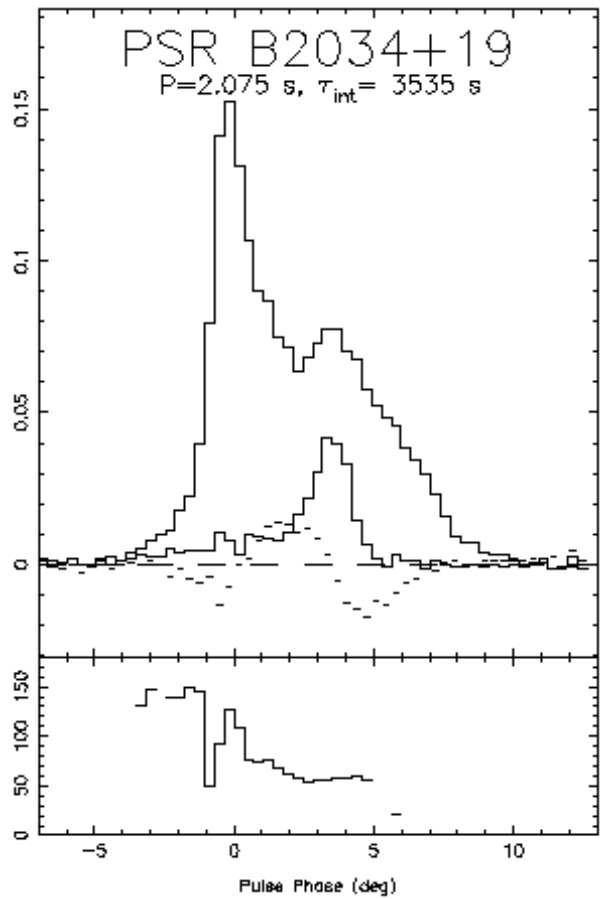}

\vspace{-3mm}

\caption{Polarization profile of PSR J2037+1942 at 1418~MHz (from
\cite{W99}). The total (higher full line), linear polarized (lower
full line) and circular polarized (dotted line) flux densities are
displayed in the upper panel. The lower panel shows the PA curve.}
  \label{fg:two_rev}
\end{figure}

\subsection{Circular Polarization of Conal-double Pulsars}

Using the polarization data of a sample of 20 conal-double pulsars
available at that time, \cite{Han98} found a strong correlation
between the sense of PA sweep and the sense of circular
polarization for conal-double pulsars, namely, a decrease of PA
accompanies with left-hand  circular polarization of conal
components, and an increase of PA with the right-hand.
Occasionally, sense reversal is observed in one cone component of
profiles.

Now, using a larger sample of 36 pulsars, the correlation is
solidly confirmed.  Table~\ref{tb:cone_double} lists all
conal-double pulsars with good measurements of circular
polarization and PA.

\begin{table}

\centering

\begin{minipage}{110mm}
\caption{Conal-double Pulsars with Significant Circular
Polarization} \label{tb:cone_double}
\end{minipage}
 \scriptsize
\tabcolsep 4mm
\begin{tabular}{llccl}
\hline\noalign{\smallskip}
PSR Name & PA  &  \multicolumn{2}{c}{Sign of $V$} & Ref. \\
            &     & Comp 1 & Comp 2 &    \\
\hline\noalign{\smallskip}
J0151$-$0635& inc & $-$ & $-$ & LM88, G98      \\
J0528$+$2200& inc & $-$ & $-$ & S84a, R83      \\
J0653$+$8051& inc &$+/-$& $-$ & G98           \\
J0754$+$3231& inc & $-$ & $-$ & R89, G98       \\
J0820$-$1350& inc & $-$ & $-$ & v97, Q95, B87   \\
J0837$+$0610& inc & $-$ & $-$ & Mc78, S84a, G98   \\
            &     &$+/-$& $-$ & S84b \\
J0959$-$4809& inc & $-$ & $-$ & H06           \\
J1015$-$5719& inc & $-$ & $-$ & H06           \\
J1110$-$5637& inc &$\cdot$& $-$ & H06         \\
J1136$+$1551& inc & $-$ & $-$ & Mc78, S84a, G98  \\
J1137$-$6700& inc & $-$ & $-$ & H06           \\
J1159$-$7910?& inc &$\cdot$& $-$ & H06         \\
J1420$-$6048& inc & $-$ & $-$ & R01           \\
J1906$+$0641& inc & $-$ & $\cdot$& W99        \\
J1915$+$1606& inc &$-/+$& $-$ & C90           \\
J1921$+$2153& inc &$-/+$& $-$ & W99           \\
J1954$+$2923& inc & $-$ & $-$ & G98             \\
J2022$+$2854& inc & $-$ & $-$ & C78, S84a, W99      \\
J2046$+$1540& inc & $-$ & $-$ & G98 \\
J2053$-$7200& inc &$+/-$ &$-$ & Q95, M98 \\
            &     &$-/+$ &$-$ & Q95, H06 \\
J2124$+$1407& inc & $-$ & $-$ & W04 \\
\hline\noalign{\smallskip}
J0055$+$5117& dec & $+$ & $+$ & G98 \\
J0304$+$1932& dec & $+$ & $+$ & R83, R89, W99\\
J0631$+$1036& dec & $+$ & $+$ & Z96 \\
J1041$-$1942& dec &$-/+$& $+$ & LM88, G98 \\
J1123$-$4844& dec & $+$ & $+$ & M98 \\
J1302$-$6350& dec & $+$ &$\cdot$& MJ95 \\
J1345$-$6115& dec &$+$&$\cdot$& H06 \\
J1527$-$3931& dec & $+$ & $+$ & M98 \\
J1731$-$4744& dec & $+$ &$\cdot$& H77, Mc78, v97\\
J1751$-$4657& dec &$-/+$& $+$ & M98 \\
J1803$-$2137& dec & $+$ & $+$ & G98 \\
J1826$-$1344& dec & $+$ & $+$ & G98 \\
J2055$+$2209& dec & $+$ & $+$ & G98 \\
J2324$-$6054& dec & $+$ &$\cdot$& Q95\\
J2346$-$0609& dec &$\cdot$&$+$& M98 \\
\hline\noalign{\smallskip} \multicolumn{4}{l}{ ?~ Not so sure for
conal-double pulsar.}
\\
\end{tabular}
\end{table}

We also checked if there is any correlation between the
polarization percentage and the maximum sweep rate of PA. Ideally,
the PA should follow the $S$-shaped curve across the pulse profile
as described by the rotating vector model (\cite{RC69}). The
maximum rate of polarization sweep, which occurs when the line of
sight passes closest to the magnetic axis, is given by
$\left(\frac{d\psi}{d\phi}\right)_m
=\frac{\sin\alpha}{\sin\beta}$, where $\psi$ is the PA, $\phi$ is
the longitude, $\alpha$ is the inclination of the magnetic axis to
the rotation axis, and $\beta$ is the impact parameter given by
$\beta=\zeta-\alpha$, where $\zeta$ is the inclination of the
observer direction to the rotation axis. The value of
$\left|\frac{d\psi}{d\phi}\right|_m$ very sensitively depends on
$\left|\beta\right|$. Smaller $\left|\beta\right|$, i.e. the
magnetic axis closer to the observer direction, gives a larger
$\left|\frac{d\psi}{d\phi}\right|_m$. Figure~2 shows the
relationship between \relv~ and
$\left(\frac{d\psi}{d\phi}\right)_m$ at 1400~MHz for conal-double
pulsars. There are only 27 pulsars in Figure~2 because some
pulsars have not been observed near 1400~MHz or the observed PA
does not have a good enough $S$-shaped curve to estimate
$\left|\frac{d\psi}{d\phi}\right|_m$. Pulsars are located in the
second and forth quadrants in Figure~2, which confirms the
correlation between the sign of PA swing and the sense of $V$.
Furthermore, we noticed that $|\langle V\rangle|/S$ tends to
decrease with $\left|\frac{d\psi}{d\phi}\right|_m$.

\begin{figure}
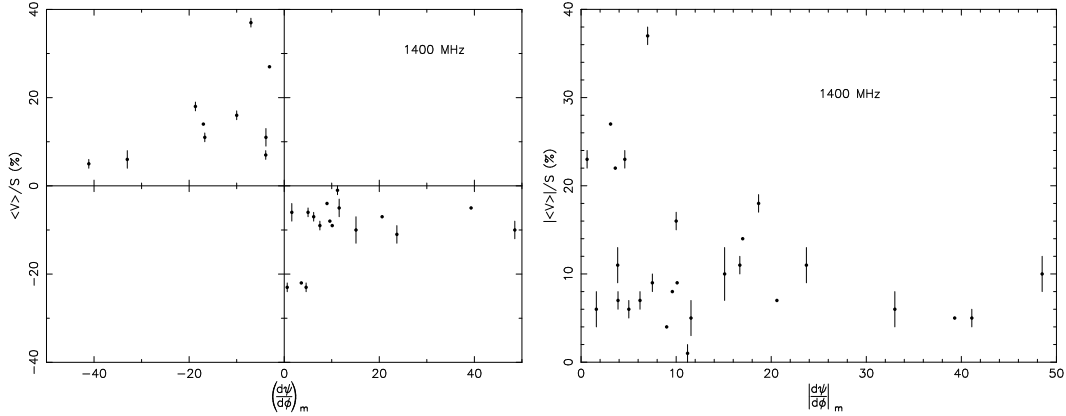


\vs
 \centering
\includegraphics[angle=-90,width=70mm]{figure2a.eps}
\includegraphics[angle=-90,width=70mm]{figure2b.eps}

  \caption{ The maximum polarization sweep rate
  $\left(\frac{d\psi}{d\phi}\right)_m$ is related to the fractional circular
  polarization \relv~ at 1400~MHz. The data of $\left(\frac{d\psi}{d\phi}\right)_m$
  were taken from Gould (1994), Han et al.\ (2006), or estimated by ourselves.}
  \label{fg:slope_cp}
\end{figure}

\subsection{Circular Polarization with Frequency}

The circular polarization of some pulsars clearly changes with
frequency. von Hoensbroech \& Lesch (1999) showed three pulsars
with a trend of increasing circular polarization with frequency,
which was interpreted in terms of propagating natural wave modes
in pulsar magnetosphere.

The variation of degree of circular polarization with frequency is
very different from pulsar to pulsar.  Figure~3 shows eight good
examples: Four of the pulsars show their circular polarizations
increasing with frequency, but the other four in the latter part
of the figure show a decrease.

In some pulsars, the sign of sense reversal clearly changes with
frequency. PSR~J2053$-$7200 
shows a sense reversal near the intersection of two components
from the left-hand to right-hand at low frequencies (\cite{Q95};
\cite{M98}; \cite{v97}), but from the right-hand to left-hand at
high frequencies (\cite{Q95}; \cite{H06}).

   \begin{figure}
   \centering

\includegraphics[angle=-90,width=135mm]{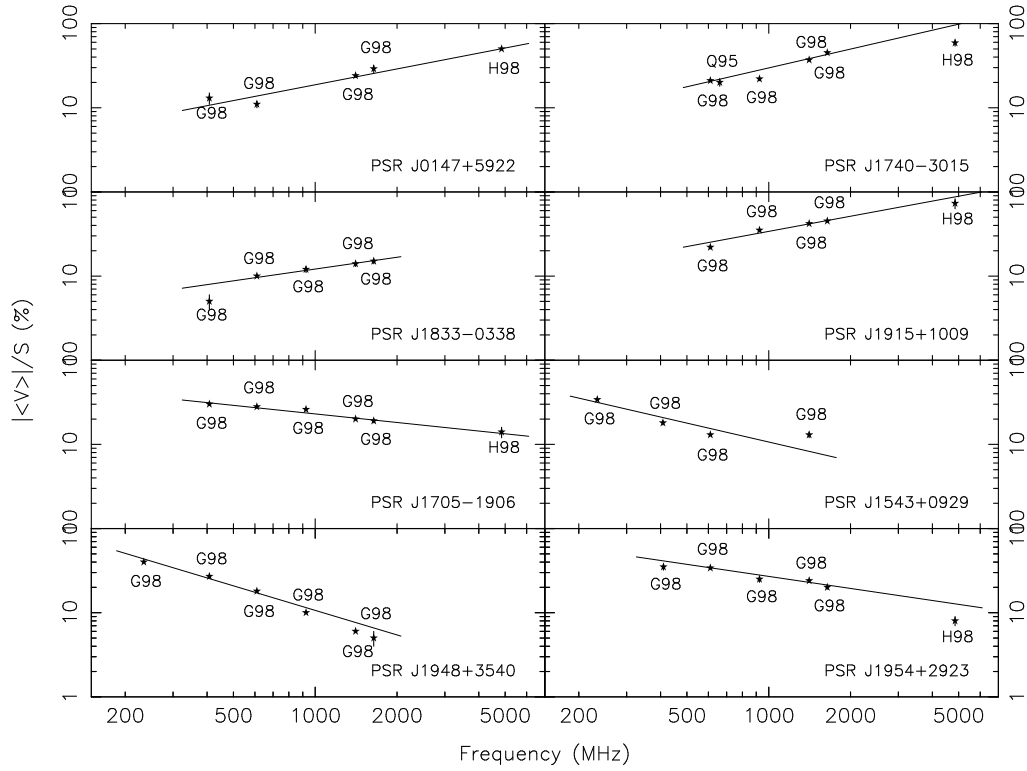}

\begin{minipage}{130mm}
\caption{Eight pulsars with clear variation of circular
polarization with frequency.}
   \label{fg:absvfreq}\end{minipage}
   \end{figure}

\subsection{Circular Polarization in Normal Pulsars and Millisecond Pulsars}

Compared to normal pulsars, millisecond pulsars have weaker
surface magnetic fields, wider profiles, and a different profile
dependence on frequency (\cite{Kramer98}; \cite{Kramer99}). Though
their PA variations are often more complicated, most of them
appear to follow the rotating vector model. The basic radio
emission mechanism may be similar for millisecond pulsars and
normal pulsars. \cite{X98} found that the fractional absolute
circular polarization is higher for millisecond pulsars than for
normal pulsars, based on observations at 1410~MHz.

\begin{table}

\centering

\begin{minipage}{120mm}
\caption{ Circular Polarization of Millisecond Pulsars Near
1400~MHz}
  \label{tb:MSP}\end{minipage}

\tabcolsep 4mm

\scriptsize
  \begin{tabular}{lrrrrcl}
  \hline\noalign{\smallskip}
PSR Name & Period & \multicolumn{1}{c}{\relmodv} & \multicolumn{1}{c}{\relv}
& \multicolumn{1}{c}{Err.}& \multicolumn{1}{c}{Freq.} & Ref. \\
  & (ms) & (\%) & (\%) & (\%) & (MHz) &  \\
  \hline\noalign{\smallskip}
J0437$-$4715 &  5.76 & 11 &$  -5 $&  1 & 1512 & N97\\
J0711$-$6830 &  5.49 & 16 &$     $&    & 1405 & O04\\
J1022$+$1001 & 16.45 & 10 &$     $&    & 1405 & O04\\
J1045$-$4509 &  7.47 & 14 &$     $&    & 1373 & O04\\
J1600$-$3053 &  3.60 &  3 &$     $&    & 1373 & O04\\
J1603$-$7202 & 14.84 & 28 &$     $&    & 1405 & O04\\
J1623$-$2631 & 11.08 & 18 &$   1 $&  3 & 1331 & MH04\\
J1629$-$6902 &  6.00 & 14 &$     $&    & 1373 & O04\\
J1643$-$1224 &  4.62 & 11 &$  -1 $&  1 & 1331 & MH04\\
J1713$+$0747 &  4.57 &  3 &$     $&    & 1373 & O04\\
J1730$-$2304 &  8.12 & 17 &$     $&    & 1405 & O04\\
J1748$-$2446 & 11.56 & 14 &$     $&    & 1414 & S99\\
J1757$-$5322 &  8.86 & 19 &$     $&    & 1373 & O04\\
J1804$-$0735 & 23.10 & 13 &$ -12 $&  1 & 1408 & G98\\
J1857$+$0943 &  5.36 &  6 &$     $&    & 1373 & O04\\
J1909$-$3744 &  2.94 & 14 &$     $&    & 1373 & O04\\
J1911$-$1114 &  3.63 & 17 &$     $&    & 1373 & O04\\
J1915$+$1606 & 59.03 & 17 &$ -10 $&  3 & 1408 & G98\\
J1933$-$6210 &  3.54 &  6 &$     $&    & 1373 & O04\\
J1939$+$2134 &  1.56 &  3 &$     $&    & 1414 & S99\\
J2051$-$0827 &  4.51 & 10 &$     $&    & 1341 & O04\\
J2124$-$3358 &  4.93 &  6 &$  -1 $&  1 & 1327 & MH04\\
J2129$-$5721 &  3.73 & 30 &$     $&    & 1373 & O04\\
J2145$-$0750 & 16.05 &  5 &$     $&    & 1373 & O04\\
  \noalign{\smallskip}\hline\\
  \end{tabular}
\end{table}

Here we compiled a sample of millisecond pulsars observed near
1400~MHz as listed in Table~\ref{tb:MSP} (\cite{S99}; \cite{MH04};
\cite{O04}) and compared their circular polarization with that of
normal pulsars. The distributions of degree of circular
polarization of millisecond pulsars and normal pulsars are
marginally different, as shown in Figure~4. The Kolmogorov-Smirnov
test returned a probability of 16.49\% for the two populations
being from the same distribution.

   \begin{figure}
   \centering

\includegraphics[angle=-90,width=100mm]{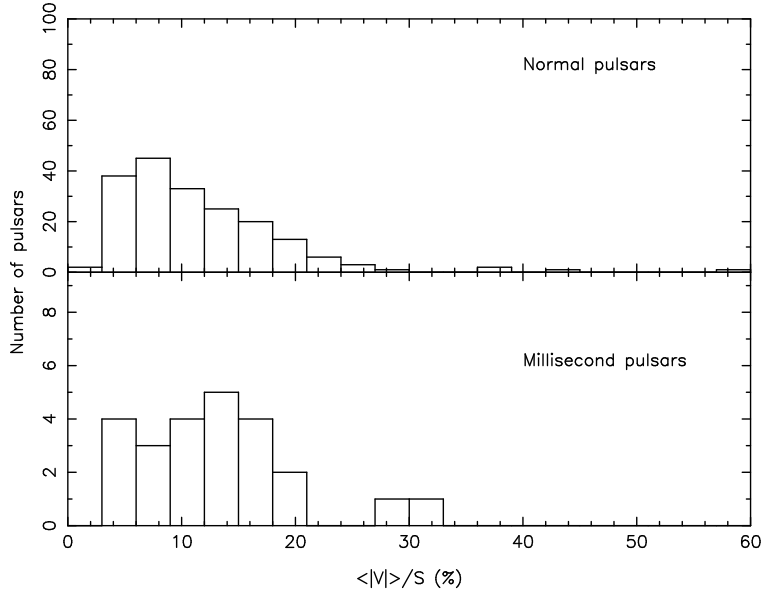}

\begin{minipage}{110mm}
\caption{Comparison of fractional absolute circular polarization
of normal pulsars and millisecond pulsars.}
   \label{fg:compa}\end{minipage}
   \end{figure}

\section{Discussion}

Circular polarization can be generated by several emission
mechanisms including curvature emission, coherent emission and
cyclotron absorption. \cite{Michel87} first noted that curvature
emission can explain the sense reversal of the circular
polarization. Following this model, \cite{GS90} re-examined
curvature radiation and demonstrated that circular polarization
could have a sense reversal near pulse center. \cite{Xu00}
considered coherent emission of a bunch of electrons with inverse
Compton scattering, and found that circular polarization can be
produced at low emission altitudes. On the other hand, cyclotron
absorption may also produce circular polarization (Melrose 2003).

   \begin{figure}
   \centering

\includegraphics[angle=-90,width=110mm]{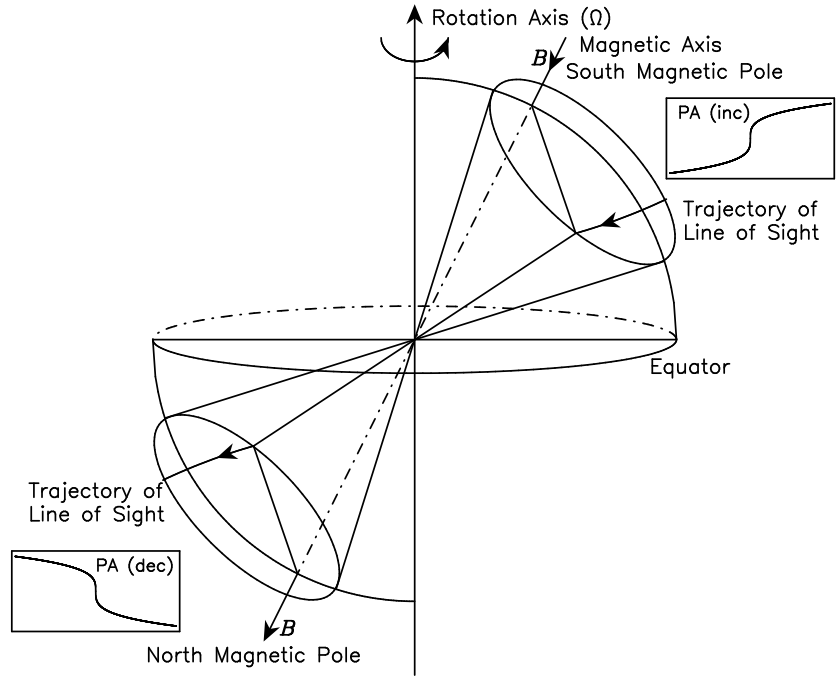}
\caption{IAG geometry of a pulsar.
The beam geometry for the inner annular gap model. The PA
decreases or increases with the pulse longitude as the line of
sight cuts the beam between the equator and the Northern or
Southern magnetic pole.
}
   \label{fg:gap}
   \end{figure}

Propagation effects can induce circular polarization or at least
influence circular polarization. There are two kinds of
propagation effect. One is in the pulsar magnetosphere and the
other in interstellar medium (ISM). \cite{PL00} investigated
refraction and polarization transferring in an ultra-relativistic
highly magnetized pulsar plasma. They found that circular
polarization arises out of rotation of the magnetosphere. Two main
types of circular polarization defined by \cite{RR90} can be
explained by refraction in the plasma with non-axisymmetric
density distribution and by magnetosphere rotation.
\cite{petrova01} also found that the change in the sense of
circular polarization can occur near the orthogonal transitions or
from non-orthogonality of the observed modes.

Macquart \& Melrose (2000) discussed a scintillation-induced
circular polarization in the interstellar medium due to rotation
measure gradient. The degree of circular polarization induced by
diffractive scintillation at lower frequency is more significant.
We calculated this effect due to rotation measure gradient
(\cite{v97}; \cite{HMQ99}; \cite{W04}) and found this effect is
very small (less than a few percent) except for a few pulsars at
low frequency.

The correlation between the sense of PA variation and the sense of
$V$ in conal-double pulsars may give some constraints to the
geometry and mechanism of pulsar emission. \cite{Qiao04} proposed
the inner annular gap (IAG) to explain the emission from pulsars.
For neutron stars, an IAG can be formed only for a pulsar
(\mbox{\boldmath $\Omega \cdot B<0$}), not for an antipulsar
(\mbox{\boldmath $\Omega \cdot B>0$}), and the beam is asymmetric
in shape, much larger toward the equator. According to this model,
as Figure~5 shows, conal-double pulses are more likely generated
in the region close to equator of the pulsar (\mbox{\boldmath
$\Omega \cdot B<0$}).
We also know that the PA decreases or increases with the pulse
longitude when the line of sight cuts the beam between the equator
and the Northern or Southern magnetic pole.
Based on the observed correlation in conal-double pulsars and the
properties of IAG model, the conal emission of pulsars in the
first part of Table~\ref{tb:cone_double} is produced from the
South magnetic pole and has right-hand circular polarizations,
whereas the emission of pulsars in the second part of
Table~\ref{tb:cone_double} is from the North magnetic pole and has
left-hand circular polarization.

\section{Conclusions}

Circular polarization in pulsars shows
diverse  patterns. Though sense reversal of circular polarization
often occurs in the core components, it can also happen in the
cone components or near the intersection between components. We
confirm the correlation between the sense of circular polarization
and the sense of position angle sweep for conal-double pulsars.
Circular polarization of some pulsars get stronger with frequency,
but others get weaker. The senses of circular polarization of
conal-double pulsars may be related to the different magnetic
poles.

\vs\vs\no {\bf Acknowledgements}~~ We thank Dick Manchester, Joel
Weisberg, Guojun Qiao, Kejia Li, Xiaohui Sun and Hui Men for their
helpful comments. The authors are supported by the National
Natural Science Foundation of China (Nos.\,10025313, 10521001 and
10473015).


 \label{lastpage}
\end{document}